\documentclass{xjkas}

\def\year{2015} % publication year
\def\month{June} % publication month
\def\volume{00} % journal volume
\def\beginpage{1} % first page of article
\def\endpage{4} % last page of article
\def\received{March 11, 2015} % date paper was received by JKAS
\def\accepted{June 2, 2015} % date of acceptance
\date{Received \received; accepted \accepted}

\usepackage{flushend} %% balance columns on last page
\newcommand\ion[2]{{#1}\,{\sc #2}} % ions: \ion{C}{iv} = C IV

\def\ao{a_{0}}

\def\mss{m\,s$^{-2}$}
\def\kmsmpc{km\,s$^{-1}$\,Mpc$^{-1}$}
\def\Md{M_{\rm dyn}}
\def\Mo{M_{0}}
\def\Mg{M_{\rm g}}
\def\Tg{T_{\rm g}}
\def\vc{v_{\rm c}}
\def\gn{g_{\rm N}}
\def\mg{m_{\rm g}}
\def\mpl{m_{\rm P}}
\def\lpl{l_{\rm P}}
\def\Ho{H_{0}}
\def\Ol{\Omega_{\Lambda}}
\def\rhog{\rho_{\rm g}}
\def\ng{n_{\rm g}}

\title{
Milgrom's Law and $\mathbf{\Lambda}$'s Shadow: How Massive Gravity\\ 
Connects Galactic and Cosmic Dynamics
}

\author[]{Sascha~Trippe}
\affil[]{Department of Physics and Astronomy, Seoul National University, Seoul 151-742, Korea; \email{trippe@astro.snu.ac.kr}}
\def\corrauthor{
S. Trippe
}

\def\runningauthor{
Trippe
}

\def\runningtitle{
How Massive Gravity Connects Galactic and Cosmic Dynamics
}

\def\keywords{
gravitation --- cosmology --- dark matter --- dark energy
}

\def\abstracttext{
Massive gravity provides a natural solution for the dark energy problem of cosmology and is also a candidate for resolving the dark matter problem. I demonstrate that, assuming reasonable scaling relations, massive gravity can provide for Milgrom's law of gravity (or ``modified Newtonian dynamics'') which is known to remove the need for particle dark matter from galactic dynamics. Milgrom's law comes with a characteristic acceleration, Milgrom's constant, which is observationally constrained to $\ao\approx1.1\times10^{-10}$\,\mss. In the derivation presented here, this constant arises naturally from the cosmologically required mass of gravitons like $\ao\propto c\sqrt{\Lambda}\propto c\Ho\sqrt{3\Ol}$, with $\Lambda$, $\Ho$, and $\Ol$ being the cosmological constant, the Hubble constant, and the third cosmological parameter, respectively. My derivation suggests that massive gravity could be the mechanism behind both, dark matter and dark energy.
}

\begin{document}
\jkashead %% set title, authors, abstract, etc.

%%%%%%%%%%%%%%%%%%%%%%%%%%%%%%%%%%%%%%%%%%%%%%%%%%%%%%%%%%%%%%%%%%%%%
%%% BEGIN MAIN TEXT HERE %%%%%%%%%%%%%%%%%%%%%%%%%%%%%%%%%%%%%%%%%%%%
%%%%%%%%%%%%%%%%%%%%%%%%%%%%%%%%%%%%%%%%%%%%%%%%%%%%%%%%%%%%%%%%%%%%%

\section{Introduction \label{sec:intro}}

Modern standard cosmology suffers from two critical issues: the \emph{dark matter} problem and the \emph{dark energy} problem. Canonical $\Lambda$CDM cosmology (e.g., \citealt{bahcall1999}) requires about 95\% (e.g., \citealt{ade2014}) of the mass/energy content of the universe to be provided by exotic dark components for which physical counterparts have not been identified. Dark matter and dark energy are commonly assumed to be unrelated: dark matter is identified with as yet undiscovered new exotic elementary particles, whereas dark energy is linked to a \emph{cosmological constant}, $\Lambda$, which is inserted into Einstein's field equations and which triggers an accelerated expansion of the universe on its largest scales.

$\Lambda$CDM cosmology is based on the assumption that gravitation is Einsteinian on all scales. Modified theories of gravity might be able to describe the universe without dark components; this idea has initiated a vast number of works exploring multiple models of gravitation (see \citealt{clifton2012} for an exhaustive review). A quantum-field theoretical modified theory of gravity, first proposed by \citet{fierz1939}, is \emph{massive gravity} in which gravitation is mediated by virtual bosons, \emph{gravitons}, that have a very small (compared to all other elementary particles) non-zero mass. Massive gravitons, being virtual exchange particles, have a limited life time governed by Heisenberg's uncertainty relation for energy and time. If the mass of the graviton is such that its life time is $\sqrt{1/\Lambda}$, the decay of gravity induces an accelerated expansion of the universe like the one actually observed -- thus providing an elegant resolution of the dark energy problem (recently, e.g., \citealt{cardone2012, clifton2012, hinterbichler2012, volkov2012, tasinato2013, defelice2013, derham2014}).

On smaller scales, the dynamics of galaxies is in excellent agreement with a modification of Newtonian dynamics (the \emph{MOND paradigm}) in the limit of small acceleration (or gravitational field strength) $g$, expressed in \emph{Milgrom's law}
\begin{equation}
\label{eq:milgrom}
\gn = \mu(g/\ao)\,g
\end{equation}
\citep{milgrom1983a} with $\gn$ being the Newtonian field strength, $\ao$ denoting \emph{Milgrom's constant}, and $\mu$ being a transition function with $\mu=1$ for $g/\ao\gg1$ and $\mu=g/\ao$ for $g/\ao\ll1$. Such a modification implies that the ratio of dynamical and luminous masses of a stellar system, the \emph{mass discrepancy} $\Md/\Mo$, exceeds unity and becomes a function of acceleration (or field strength) when assuming Newton's law of gravity. In the ``deep MOND'' limit $g/\ao\ll1$, the circular speed $\vc$ of stars in rotation-supported dynamical systems (especially disk galaxies) is
\begin{equation}
\label{eq:vc}
\vc^4 = G\,\Mo\,\ao
\end{equation}
with $G$ being Newton's constant \citep{milgrom1983a, milgrom1983b, milgrom1983c}; note that $g=\vc^2/r=G\Md/r^2$ and $\gn=G\Mo/r^2$ for circular orbits with radius $r$. For pressure-supported systems (especially elliptical galaxies and galaxy clusters) with 3D velocity dispersion $\sigma$, the mass--velocity relation takes the form
\begin{equation}
\label{eq:sigma}
\sigma^4 = \frac{4}{9}\,G\,\Mo\,\ao
\end{equation}
\citep{milgrom1984,milgrom1994}. Equations~(\ref{eq:milgrom}, \ref{eq:vc}, \ref{eq:sigma}), with $\ao\approx1.1\times10^{-10}$\,\mss, naturally provide for the fundamental scaling laws of galactic kinematics, specifically the baryonic Tully--Fisher, baryonic Faber--Jackson, mass discrepancy--acceleration (MDA), and surface mass density--acceleration relations, plus the asymptotic flattening of rotation curves and the occurrence of ``dark rings'' in galaxy clusters \citep{sanders1994, sanders2010, rhee2004a, rhee2004b, mcgaugh2004, mcgaugh2005a, mcgaugh2005b, milgrom2008, mcgaugh2011, gentile2011, cardone2011, famaey2012, trippe2014, walker2014, wu2015, milgrom2015, chae2015}. Milgrom's law eliminates particle dark matter from galactic dynamics together with its substantial difficulties (cf., e.g., \citealt{kroupa2012, kroupa2015}).

As pointed out by \citet{trippe2013a, trippe2013b, trippe2013c,trippe2015}, massive gravity can, at least in principle, provide Milgrom's law together with an expression for the function $\mu$ in agreement with observations. Following up on this ansatz, I investigate the link of galactic dynamics to the cosmic expansion history. A physical connection is established by the graviton mass $\mg\propto\sqrt{\Lambda}$. Milgrom's constant is given by $\ao\propto c\sqrt{\Lambda}\propto c\Ho\sqrt{3\Ol}$, with $c$, $\Lambda$, $\Ho$, and $\Ol$ being the speed of light, the cosmological constant, the (present-day) Hubble constant, and the (present-day) third cosmological parameter, respectively. Accordingly, massive gravity might offer a natural alternative explanation for the phenomena conventionally associated with dark matter and dark energy.

\section{Calculations \label{sec:calc}}

Massive gravity implies \citep{trippe2013a, trippe2013b, trippe2013c,trippe2015} that any luminous (``baryonic'') mass $\Mo$ is the source of a spherical graviton halo with mass density
\begin{equation}
\label{eq:rho}
\rhog(R) = \mg\,\ng(R) = \mg\,\eta\,\Mo\,R^{-2}
\end{equation}
where $\mg$ is the graviton mass, $\ng$ is the graviton particle density, $R$ is the radial coordinate, and $\eta$ is a scaling factor. The proportionality $\rhog\propto\Mo R^{-2}$ follows from (i) consistency with the classical force law in the Newtonian limit, and (ii) the inverse-square-of-distance law of flux conservation (but see also the discussion in Section~\ref{sec:discuss}). For a circular orbit of radius $r$ around $\Mo$, the total, dynamical mass experienced by a test particle follows from integrating $\rhog(R)$ from 0 to $r$,
\begin{equation}
\label{eq:mdyn}
\Md = \Mo + \Mg = \Mo\left(1 + 4\,\pi\,\eta\,\mg\,r\right)
\end{equation}
with $\Mg$ being the mass contributed by the graviton halo. One can bring this expression into the more intuitive form
\begin{equation}
\label{eq:md01}
\frac{\Md}{\Mo} = 1 + 4\,\pi\,f\,\frac{\mg}{x}\,\frac{r}{y}
\end{equation}
where $f$ is a dimensionless number, $x$ is a mass scale, and $y$ is a length scale (with $\eta\equiv f/(xy)$).

With gravitons being virtual exchange particles, their effective mass can be estimated from (cf., e.g., \citealt{griffiths2008}) Heisenberg's uncertainty relation for energy and time,
\begin{equation}
\label{eq:heisenberg}
\mg c^2\times\Tg\approx\hbar
\end{equation}
with $\Tg$ being the life time of gravitons and $\hbar$ being the reduced form of Planck's constant. Consistency with the cosmic expansion history requires $\Tg\approx\sqrt{1/\Lambda}$, resulting in
\begin{equation}
\label{eq:mg01}
\mg \approx \frac{\hbar}{c^2}\,\sqrt{\Lambda} ~ .
\end{equation}
According to the standard ``cosmic triangle'' formalism (e.g., \citealt{bahcall1999}), $\Lambda=3\Ho^2\Ol$, resulting in
\begin{equation}
\label{eq:mg02}
\mg \approx \frac{\hbar}{c^2}\,\Ho\,\sqrt{3\Ol} ~ ;
\end{equation}
for $\Ho\approx70$\,\kmsmpc\ and $\Ol\approx0.7$ (e.g., \citealt{ade2014}), $\Lambda\approx1.1\times10^{-35}$\,s$^{-2}$ and thus $\mg\approx4\times10^{-69}\,{\rm kg}\approx2\times10^{-33}$\,eV\,$c^{-2}$.

A natural choice for a quantum-physical mass scale, $x$ in our case, is the \emph{Planck mass}
\begin{equation}
\label{eq:mplanck}
x \equiv \mpl = \sqrt{\frac{\hbar\,c}{G}}
\end{equation}
with $G$ being Newton's constant; combination of Equations (\ref{eq:mg02}) and (\ref{eq:mplanck}) leads to $\mg/\mpl\approx2\times10^{-61}$. The choice for the scale $y$ is less obvious; for our purpose, we require a scale characteristic for gravitationally bound dynamical systems. A good choice (to be discussed in Section~\ref{sec:discuss}) is the magnitude of the gravitational potential,
\begin{equation}
\label{eq:epsilon}
\epsilon = \frac{G\,\Md}{r\,c^2}
\end{equation}
(cf. also \citealt{baker2015}). As $\epsilon$ is dimensionless, we can supply $y$ with the unit of a length by multiplying $\epsilon$ and the \emph{Planck length}
\begin{equation}
\label{eq:lplanck}
\lpl = \sqrt{\frac{\hbar\,G}{c^3}}
\end{equation}
so that $y\equiv\epsilon\,\lpl$. 

Inserting Equations~(\ref{eq:mg01}, \ref{eq:mplanck}, \ref{eq:epsilon}, \ref{eq:lplanck}) into Equation~(\ref{eq:md01}) leads to
\begin{equation}
\label{eq:md02}
\frac{\Md}{\Mo} = 1 + 4\,\pi\,f\,c\,\sqrt{\Lambda}\ \frac{r^2}{G\,\Md} ~ .
\end{equation}
Using the classical expression for gravitational field strength, $g=G\Md/r^2$, one finds
\begin{equation}
\label{eq:md03}
\frac{\Md}{\Mo} = 1 + 4\,\pi\,f\ \frac{c\,\sqrt{\Lambda}}{g}
\end{equation}
or, using the expression given by Equation~(\ref{eq:mg02}),
\begin{equation}
\label{eq:md04}
\frac{\Md}{\Mo} = 1 + 4\,\pi\,f\ \frac{c\,\Ho\,\sqrt{3\,\Ol}}{g} ~ .
\end{equation}
Comparison of Equations~(\ref{eq:md03}, \ref{eq:md04}) to the expression following from the ``simple $\mu$ function'' (e.g., \citealt{famaey2012}) of MOND,
\begin{equation}
\label{eq:mu}
\frac{\Md}{\Mo} = 1 + \frac{\ao}{g} ~ ,
\end{equation}
shows that the expressions are equivalent, with Milgrom's constant being given by $\ao=4\pi f c \sqrt{\Lambda}$. Matching this expression with the empirical value $\ao\approx1.1\times10^{-10}$\,\mss\ requires a (universal) value of $f\approx0.0088$ (or $4\pi f \approx 1/9$). It is straightforward to see that in the limit $g\ll\ao$, Equation~(\ref{eq:mu}) leads to Equation~(\ref{eq:vc}).

\section{Discussion \label{sec:discuss}}

As demonstrated in Section~\ref{sec:calc}, massive gravity can be used to link the cosmological constant to a specific version of Milgrom's law, namely the one comprising the ``simple $\mu$ function''. The transition function given by Equation~(\ref{eq:mu}), with $\ao=(1.06\pm0.05)\times10^{-10}$\,\mss, is in excellent agreement with the observed mass discrepancy--acceleration relation of disk galaxies \citep{trippe2013c}. I thus present a physical mechanism that provides astrophysically meaningful expressions for both $\mu$ and Milgrom's constant. My derivation implies that dark matter and dark energy could be understood as phenomena arising from the same effect -- the non-zero mass and thus finite life time of gravitons. This mechanism also explains why $\ao\sim c\Ho$ as already noted by \citet{milgrom1983a}. I also note that a similar approach has been discussed by \citet{vanputten2014} in the specific context of de-Sitter space \citep[but see also][]{vanputten2015}.

The result expressed in Equation~(\ref{eq:md03}) critically depends on the choice of the scales $x$ and $y$. Identifying $x$ with the Planck mass is a natural choice. Identification of $y$ with the magnitude of the gravitational potential, $\epsilon$, (multiplied with the Planck length) is suggested by the properties of gravity on astronomical scales. As pointed out recently by \citet{baker2015}, astrophysical gravitating systems are characterized completely by their location in a two-dimensional plane spanned by (i) the parameter $\epsilon$ and (ii) the \emph{Kretschmann scalar} which corresponds to the magnitude of the Riemann curvature tensor (thus quantifying the local strength of spacetime curvature in the frame of general relativity) and which takes the form  $\xi=\sqrt{48}\,G\Md/(r^3c^2)$ for spherical systems. However, $\xi$ can be expressed in units of the cosmic curvature, $\Lambda/c^2$; thus the curvature (or at least a specific value thereof) already appears in Equation~(\ref{eq:mg01}), which suggests the introduction of $\epsilon$ (multiplied with the Planck length for dimensional reasons) as a new scaling parameter. Conveniently, my choice of $x$ and $y$ cancels out the constant $\hbar$ coming with the graviton mass (Equation~(\ref{eq:mg01})) because $\mpl\,\lpl=\hbar/c$. The dimensionless factor $f$ arises from the need to provide a normalization factor for the graviton halo density in Equation~(\ref{eq:rho}) of the form $\eta\equiv f/(xy)$; I emphasize that $f$ is the only free parameter left in the derivation because neither $\mpl$ nor $\lpl$ nor $\epsilon$ are tunable. For mass density distributions, scaling factors $0<f\leq1$ -- in our case, $f\approx0.9$\% -- can be interpreted generically as filling or efficiency factors (in the absence of further information, as is the case here). I also note that Equation~(\ref{eq:rho}) follows explicitly from the limiting case of Newtonian gravity. Obviously, the inverse-square-of-distance law of gravity does not hold globally in massive gravity but only in the limit $g\gg\ao$. The choice of Equation~(\ref{eq:rho}) is motivated by the need to connect Newtonian and modified Newtonian gravity in the strong field limit.

Even though they are quite intuitive already, the relations presented in Section~\ref{sec:calc} should eventually follow from, or be included in, a complete theory of massive gravity -- which has not been found yet. Especially, it will be necessary to connect the classical derivation of Equation~(\ref{eq:mu}) with general relativity and the resulting deviations from Newtonian gravity in the strong-field limit ($g\gg\ao$). A complete theory should also provide for the basic properties of gravitons, especially the absence of graviton--graviton interactions which is required for the validity of the classical force law (as used in Equation~(\ref{eq:rho}); cf., \citealt{trippe2013a} -- but see also \citealt{vainshtein1972, babichev2013, avilez2015}). Last but not least, a complete theory of massive gravity should also comprise non-dynamical effects that are currently unexplained in the frame of general relativity; possible candidates are the ``radio flux ratio anomaly'' affecting the images of multiply gravitationally lensed quasars (e.g., \citealt{xu2015}, and references therein) and the discrepancy between mass estimates based on galactic dynamics and those based on gravitational lensing found recently for filaments of the Virgo cluster of galaxies \citep{lee2015}.

\section{Summary and Conclusions \label{sec:conclude}}

I explore a possible physical connection between dark matter and dark energy. Assuming that gravitation is mediated by virtual gravitons with non-zero mass (massive gravity), the resulting limited life time of gravitons provides for a decay of gravity on cosmological scales and thus an accelerated expansion of the universe (``dark energy''); the graviton mass follows from the cosmological constant like $\mg\approx\hbar\sqrt{\Lambda}/c^2$ via Heisenberg's uncertainty relation. Massive gravity also implies that any luminous (``baryonic'') mass is surrounded by a (electromagnetically invisible) halo of gravitons that contributes additional mass (``dark matter''). Assuming reasonable scaling relations for such graviton halos, one recovers Milgrom's law of gravity and finds that Milgrom's constant is $\ao = 4\pi f c \sqrt{\Lambda} = 4\pi f c\Ho\sqrt{3\Ol}$ with $f\approx0.9\%$ (or $4\pi f\approx1/9$).

My derivation suggests that dark matter and dark energy could be interpreted as two effects arising from the same physical mechanism: massive gravity. This would imply a natural connection between the dynamics of galaxies and the dynamics of the universe.

%%% ACKNOWLEDGMENTS (IF ANY) %%%%%%%%%%%%%%%%%%%%%%%%%%%%%%%%%%%%%%%%

\acknowledgments

I am grateful to {\small\sc Benoit Famaey} and {\small\sc Christian Boily} (both: Observatoire Astronomique de Strasbourg) for inspiring discussion, and to an anonymous referee for valuable comments. I acknowledge financial support from the Korean National Research Foundation (NRF) via Basic Research Grant 2012-R1A1A-2041387.

%%% APPENDICES (IF ANY) %%%%%%%%%%%%%%%%%%%%%%%%%%%%%%%%%%%%%%%%%%%%%

%%% CALL LIST OF REFERENCES (natbib STYLE) %%%%%%%%%%%%%%%%%%%%%%%%%%

\end{document}